\magnification=1200
\vsize=7.5in
\hsize=5in
\tolerance 10000
\pageno=1

\def\half{{1\over 2}}
\def\fo{\vert f_1 \rangle}
\def\ft{\vert f_2 \rangle}
\def\eo{\vert e_1 \rangle}
\def\et{\vert e_2 \rangle}
\def\phase{e^{i\phi_{AB}}}
\def\m{{\bf \mu}}
\def\A{{\bf A}}
\def\r{{\bf r}}
\def\R{{\bf R}}
\def\v{{\bf v}}
\def\V{{\bf V}}
\baselineskip 12pt plus 2pt minus 2pt
\centerline{\bf \quad THE AHARONOV-CASHER EFFECT}
\bigskip
In 1984, 25 years after the prediction of the $\rightarrow$Aharonov-Bohm (AB) effect,
Aharonov and Casher [1] predicted a ``dual" effect.  In both effects, a particle is
excluded from a tubular region of space, but otherwise no force acts on it. Yet it
acquires a measurable quantum phase that depends on what is inside the tube of space
from which it is excluded.  In the AB effect, the particle is charged and the tube
contains a magnetic flux. In the Aharonov-Casher (AC) effect, the particle is neutral,
but has a magnetic moment, and the tube contains a line of charge.  Experiments in
neutron [2], vortex [3], atom [4], and electron [5] interferometry bear out the
prediction of Aharonov and Casher.  Here we briefly explain the logic of the AC effect
and how it is dual to the AB effect.

We begin with a two-dimensional version of the AB effect.  Figure 1 shows an electron
moving in a plane, and also a ``fluxon", i.e. a small region of magnetic flux (pointing
out of the plane) from which the electron is excluded.  In Fig. 1 the fluxon is in a
quantum superposition of two positions, and the electron diffracts around one of the
positions but not the other.  Initially, the fluxon and electron are in a product state
$\vert \Psi_{in} \rangle$:
$$
\vert \Psi_{in} \rangle = \half (\fo + \ft)\otimes (\eo + \et)~~~,
$$
where $\fo$ and $\ft$ represent the two fluxon wave packets and $\eo$ and $\et$ represent
the two electron wave packets.  After the electron passes the fluxon, their state $\vert
\Psi_{fin} \rangle$ is not a product state; the relative phase between $\eo$ and $\et$
depends on the fluxon position:
$$
\vert \Psi_{fin} \rangle = \half \fo \otimes (\eo + \et) + \half
\ft\otimes  (\eo + \phase \et)~~~~.
$$
Here $\phi_{AB}$ is the Aharonov-Bohm phase, and $\ft$ represents the fluxon positioned
between the two electron wave packets. Now if we always measure the position of the fluxon
and the relative phase of the electron, we discover the Aharonov-Bohm effect: the electron
acquires the relative phase $\phi_{AB}$ if and only if the fluxon lies between the two
electron paths.  But we can rewrite $\vert \Psi_{fin} \rangle$ as follows:
$$
\vert \Psi_{fin}\rangle
= \half (\fo +\ft)\otimes \eo + \half (\fo +\phase\ft) \otimes\et~~~~.
$$
This rewriting implies that if we always measure the relative phase of the {\it fluxon}
and the position of the {\it electron}, we discover an effect that is analogous to the
Aharonov-Bohm effect:  the {\it fluxon} acquires the relative phase $\phi_{AB}$ if and
only if the {\it electron} passes between the two fluxon wave packets. Indeed, the effects
are equivalent:  we can choose a reference frame in which the fluxon passes by the
stationary electron. Then we find the same relative phase whether the electron paths
enclose the fluxon or the fluxon paths enclose the electron.

In two dimensions, the two effects are equivalent, but there are two inequivalent ways
to go from two to three dimensions while preserving the topology (of paths of one
particle that enclose the other):  either the electron remains a particle and the
fluxon becomes a tube of flux, or the fluxon remains a particle (a neutral particle
with a magnetic moment) and the electron becomes a tube of charge.  These two
inequivalent ways correspond to the AB and AC effects, respectively. They are not
equivalent but {\it dual}, i.e. equivalent up to interchange of electric charge and
magnetic flux.

In the AB effect, the electron does not cross through a magnetic field; in the AC
effect, the neutral particle does cross through an electric field.  However, there is
no force on either particle.  The proof [6] is surprisingly subtle and holds only if
the line of charge is straight and parallel to the magnetic moment of the neutral
particle [8].  Hence only for such a line of charge are the AB and AC effects dual.

Duality has another derivation.  To derive their effect, Aharonov and \hbox{Casher [1]}
first obtained the nonrelativistic Lagrangian for a neutral particle of magnetic moment
$\m$ interacting with a particle of charge $e$. In Gaussian units, it is
$$
L=\half mv^2 +\half MV^2 +{e\over c} \A (\r -\R )\cdot (\v -\V )~~~,
$$
where $M, \R ,\V$ and $m ,\r ,\v$ are the mass, position and velocity of the neutral
and charged particle, respectively, and the vector potential $\A (\r-\R)$ is
$$
 \A (\r-\R) = {{\mu \times (\r -\R )}\over {\vert \r -\R \vert^3}}~~~~.
$$
Note $L$ is invariant under respective interchange of $\r ,\v$ and $\R ,\V$.  Thus $L$
is the same whether an electron interacts with a line of magnetic moments (AB effect)
or a magnetic moment interacts with a line of electrons (AC effect). However, if we
begin with the AC effect and replace the magnetic moment with an electron, and all the
electrons with the original magnetic moment, we end up with magnetic moments that all
point in the same direction, i.e. with a straight line of magnetic flux. Hence the
original line of electrons must have been straight. We see intuitively that the effects
are dual only for a straight line of charge [7].
\bigskip

\centerline{\bf \quad References}
\medskip
\noindent [Primary]

[1] {Y. Aharonov and A. Casher, ``Topological quantum effects for neutral particles",
{\it Phys. Rev. Lett.} {\bf 53}, 319-321 (1984).}

[2] {A. Cimmino, G. I. Opat, A. G. Klein, H. Kaiser, S. A. Werner, M. Arif and R.
Clothier, ``Observation of the topological Aharonov-Casher phase shift by neutron
interferometry", {\it Phys. Rev. Lett.} {\bf 63}, 380-383 (1989).}

[3] {W. J. Elion, J. J. Wachters, L. L. Sohn and J. D. Mooij, ``Observation of the
Aharonov-Casher effect for vortices in Josephson-junction arrays", {\it Phys. Rev.
Lett.} {\bf 71}, 2311-2314 (1993).}

[4] {K. Sangster, E. A. Hinds, S. M. Barnett and E. Riis, ``Measurement of the
Aharonov-Casher phase in an atomic system", {\it Phys. Rev. Lett.} {\bf 71}, 3641-3644
(1993); S. Yanagimachi, M. Kajiro, M. Machiya and A. Morinaga, ``Direct measurement of
the Aharonov-Casher phase and tensor Stark polarizability using a calcium atomic
polarization interferometer", {\it Phys. Rev.} {\bf A65}, 042104 (2002).}

[5] {M. K\"onig et al., ``Direct Observation of the Aharonov-Casher Phase", {\it Phys.
Rev. Lett.} {\bf 96}, 076804 (2006).}

[6]  {Y. Aharonov, P. Pearle and L. Vaidman, ``Comment on 'Proposed Aharonov-Casher
effect: Another example of an Aharonov-Bohm effect arising from a classical lag'", {\it
Phys. Rev.} {\bf A37}, 4052-4055 (1988).}

[7]  {I thank Prof. Aharonov for a conversation on this point.}
\medskip
\noindent [Secondary]

[8]  {For a review, see L. Vaidman, ``Torque and force on a magnetic dipole", {\it Am.
J. Phys.} {\bf 58}, 978-983 (1990).}

\bigskip

\centerline{\bf \quad Figure Caption}
\medskip
Fig. 1.  An electron and a fluxon, each in a superposition of two wave packets; the
electron wave packets enclose only one of the fluxon wave packets.

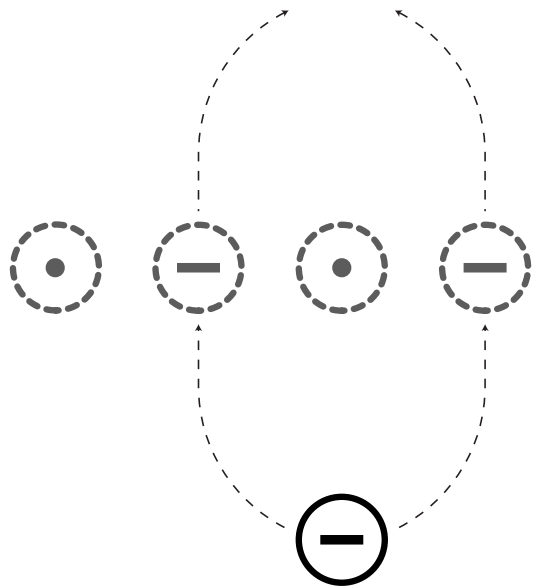
\end